\documentclass[journal,12pt, onecolumn]{IEEEtran}

\usepackage{amssymb}
\usepackage{lineno}
\usepackage{todonotes}
\usepackage{url}

\usepackage{breakurl}
\usepackage[breaklinks]{hyperref}
\usepackage{varioref}
\usepackage{xfrac}
\usepackage{mathtools}
\usepackage{doi}
\usepackage[numbers]{natbib}
\usepackage{listings}

\lstdefinestyle{mystyle}{
  basicstyle=\small\ttfamily,
  breakatwhitespace=false,         
  breaklines=true,                 
  captionpos=b,                    
  keepspaces=true,                 
  numbersep=5pt,                  
  showspaces=false,                
  showstringspaces=false,
  showtabs=false,                  
  tabsize=2
}

\begin{document}

\title{Proof of file access in a private P2P network \\ using blockchain}

\author{
        Uwe Roth\\
        Luxembourg Institute of Science and Technology \\
        ITIS Department - Security and Privacy Group \\
        5, avenue des Hauts-Fourneaux\\
        L-4362 Esch-sur-Alzette, Luxembourg\\
        uwe.roth@list.lu}
        
\maketitle

\begin{abstract}
While sharing files in a peer-to-peer (P2P) system significantly increases both the speed of retrieving the contents and the robustness of the system, tracing the access of files is not straightforward, even in the case of private P2P networks. In fact, a participant that has uploaded a file to a P2P network is not necessarily involved in its download. Additionally, due to the nature of the P2P network it is possible for a participant to already have all the fragments of a file, even before requesting it.

This work tries to address the problem of tracing file access in a private P2P file sharing network through the use of blockchains to improve quality of service and auditability. To this end, the proposed solution combines three elements: (1) A distributed hash table network that is used to distribute encrypted files with redundancy amongst the partner peers; (2) Shamir's secret sharing scheme to split the secret keys of each file; (3) A blockchain network to distribute and manage the secret shares amongst the partner peers. In fact, the latter makes access to a file undeniable to every node of the network.

The solution is relevant for consortia that manage a shared data pool on base of P2P technology with unrestricted access to files but where access to a file has to be recorded due to legal or billing reasons.

\end{abstract}

\begin{IEEEkeywords}
Shared data, Blockchain, Distributed hash table, Peer-to-peer network, Secret sharing, Transparency of access.
\end{IEEEkeywords}




\section{Introduction}
\label{sec:intoduction}
Modern file distribution and exchange solutions on base of peer-to-peer (P2Ps) networks allow quick access and fast download of files that are spread around the peers of that network. This is usually ensured by splitting a file into chunks and using redundant storage for each chunk at various peers. Then, a client may download several chunks from different peers in parallel to reduce the overall download time. At the same time, the involved peers only have to provide a fraction of the entire file.

Such P2P based solutions work quite efficiently in groups of users with unrestricted access to the files. However, since accessing a file may imply an initiation of a process, even in these cases, it is of interest and a good indicator of quality of service to determine and trace who exactly was accessing a file. Specific use case scenarios will be discussed later in Section \ref{sec:usecase}. Gaining the desired traceability is not only a problem due to the P2P nature of the file downloads which makes the concrete access of a chunk invisible to the non-involved peers. Chunks of files also might accumulate over time at a peer due to the re-organisation process of the P2P network during joining and leaving of peers. Therefore, a peer might already have all the chunks of a file without the need to request and download any further chunks. No interaction with others peers is needed that could be used to log the file access. Thus, file access traceability is rendered impossible.

An easy way to overcome this problem is the introduction of cryptographic primitives to encrypt the files before distributing in chunks. This solves the problem of having plain-text access at a peer to randomly accumulated files or to fractions of a file. But this approach introduces the problem of key-management. More precisely: Who manages the encryption keys and how can they be accessed. Again, an easy way to solve that would be the management of the keys either at a central server or at the peer that initially uploaded the file to the P2P network. In this regard, every key request can be considered as a proof that a file was accessed in plain-text, regardless of how the chunks have been spread to the P2P network. Unfortunately, this solution implies single points of failure and contradicts the initial idea of a P2P network of equal peers.

To this end, this work introduces a solution that manages to avoid single points of failures, maintains the P2P nature of the file-exchange platform and still provides evidence for the access of a file by a peer. Moreover, it adds an additional layer of a blockchain-based message exchange service which supports key management and distribution of administrative information as well as a file exchange service to a) distribute encrypted files and b) to reconstruct these files, both to ensure the desired property. It should be noted that part of the concept described in this work has already been filed as patent. The interested reader may refer to \cite{Roth2018} for more details.

The rest of this article is structured as follows: Related work and similar approaches will be presented in the next section (Section \ref{sec:relatedwork}). Potential use case scenarios will be discussed in Section \ref{sec:usecase} to give a better understanding on how the proposed solution could be used and what alternative use cases exist beside the sharing of files. Appear from the alternatives, this article focuses on the use case of a shared data pool, that will be discussed in Section \ref{sec:shareddatapool}. This section details the big picture of the solution and explains all relevant elements of the file exchange. This includes the file encryption and decryption, file distribution and download as well as the key management via a blockchain based message exchange service. The article continues in Section \ref{sec:security} with discussions about security aspects and ends with a conclusion in Section \ref{sec:conclusion}.

\section{Related work}
\label{sec:relatedwork}
The proposed solution combines a block-based file distribution network of encrypted files with a key management solution on base of secret shares and a blockchain as a communication layer. Several solutions already exist that combine some of these elements or which solve similar aspects in a different way. This section will try to classify these solutions, discuss the differences and provide some example references.

\subsection{Secret sharing and distributed storage and archiving}
\label{sec:distributedhashtableplussecretsharing}
There are already solutions available that combine the storage of encrypted files inside a distributed hash table network with the distribution of secret shares without a central controller. Some solutions are doing this with the use of a second distributed hash table as part of a friend-to-friend network \cite{Kasza2015,Marceau2005}. Focus is the management of the access to the plain-text files, but persistent logging of the access of files is not of interest in these solutions.

Alternatively the secret shares are used to be able to revert the availability of a file in future \cite{Geambasu2009}. This is similar to the proposed solution, but it is not seen as the main feature of the solution.

Another archiving solution creates the secret shares on base of the data itself and not of the encryption key \cite{Chaitanya2010}. The redundancy of a file in the network is done by creating an \textit{(m, n)} secret share schema out of each file, where \textit{n-m} nodes are able to fail. This solution uses a strategy in case of a change of network size \cite{Desmedt1997} without disclosing the files. The price that has to be paid for that strategy is, that the new \textit{(m', n')} shares have to be created out of the \textit{(m, n)} shares, by exchanging these shares in a certain way, creating a lot of traffic, as this has to be done for all archived files.

This is quite a huge difference to the proposed solution, where all files and their distributed chunks stay untouched. Instead of using the \textit{(m, n)} secret sharing scheme for the files, a similar approach is used for the encryption keys. Only they have to be re-calculated and distributed for all files.

Even if the storage of files in the shared data pool has a long-lasting aspect, the system is mainly not intended as a solution for the archiving of data, especially since the files are not provided from one system only, but potentially from all partners in the pool.

\subsection{Blockchain based digital content right management and content distribution}
\label{sec:digitalcontentrightmanagement}
The proposed solution does not provide a content right management system, but there is a very strong link to this type of solutions.

Existing solutions for digital content management on base of blockchain manage access rights and permissions \cite{Kishigami2015a}, validate or manage licenses \cite{Herbert2015,COALAIP2016}, or record the ownership \cite{McConaghy2005} on the blockchain.

The enforcement on base of that information to permit the access to files has to be done outside the blockchain and requires specific applications at the client side. Especially the integration of these components in third party processing workflows might be impossible.

In contrast to these systems, managing and accessing the secret shares to allow the decryption of a file is not a matter of authentication or authorisation at a central system or the system of the provider of the file. Also, the blockchain does not manage licenses or ownership.

Blockchain-based solutions are not yet used to communicate and persistently log messages that are part of a request-response protocol or which are used to initiate remote commands. Known solutions only use the blockchain to distribute properties and policies.

\subsection{Blockchain-based decentralised storage}
Blockchain is seen more and more as a decentralised ledger with the property to store information forever. This includes the use as a time-stamping service for data that is stored outside the blockchain \cite{StoreJLabs2018, Wilkinson2014} as well as a persistent storage for any type of information including files, assets but also meta data or policy information \cite{McConaghy2016}. This allows to create a storage marketplace \cite{Tron2016}, but is not used to share data with others.

In contrast to these technologies, the proposed solution does not provide full read-write access to distributed files. In fact, files can be written but not changed. Deletion of files is currently not foreseen but making the files unreadable can be enforced by the revocation of the encryption key, which is needed to decrypt a file.

\section{Use case scenarios}
\label{sec:usecase}

The shared data pool use case initially was developed out of concept for a research consortium that is doing research on medical data. Some dedicated members of such a consortium, known as collecting sites, are collecting specimen of donors (tissue, blood, urine) and extract medical data out of that, e.g., genomics data. In the consortium the collection site allows others to access that data and vice versa. The main reason for such a collaboration is the need for a sufficiently high number of samples, especially in case of rare deceases. Without the collaboration and exchange of data inside the consortium, not sufficient data is available to make statistically relevant claims.

What is clear in such a setup: If a set of data was accessed by an external partner, it will be used without further restriction at the site of the partner on as many computers that are needed to conduct the research. Additional systems to restrict or track the access of the data cannot be enforced by the source collecting site, mainly because of the highly customised analytic pipelines that always need unrestricted plain-text access. This does not mean that the consortium partners that are using the data are entirely free to arbitrarily use the data. Legal requirements and best practices have to be enforced, but this will happen without the involvement of the source of the data. For transparency, it is sufficient to know which specific files have been accessed by the partner in the past and which not. This might be important in case a donor of the medical data asks the collection site to withdraw the consent, as required by the GDPR \cite{gdpr,politou2018forgetting} and needs to know which third party has accessed her data.

An alternative use case is a shared consortium to exchange news. Every partner provides news to a shared pool but the access to a certain news leads to a payment obligation. Due to the fact that news is getting more and more outdated the more time has passed, only the first access by a partner is relevant for the creator of the news. Surely, such a system requires free access to meta data of a news on which base decision for downloading are taken. This is not different to the research consortium case, that also requires meaningful meta data.  

This article describes the access to files. With the same approach, also the access to an entry in a database could be managed. Instead of the use of a distributed file sharing network that distributes the chunks of a file, data is exchanged on base of a synchronised database amongst all partners. Each partner then accesses their local copy of the database to upload and read the data and uses the same scheme of encryption, decryption and key management as in the file-sharing case.

\section{Shared data pool}
\label{sec:shareddatapool}
As explained in the previous section, this article focuses on the use case of a shared data pool and not on a synchronised database.

\subsection{The big picture}
\label{sec:thebigpicture}

\begin{figure}[!htb]
    \centering
    \includegraphics[width=0.7\textwidth]{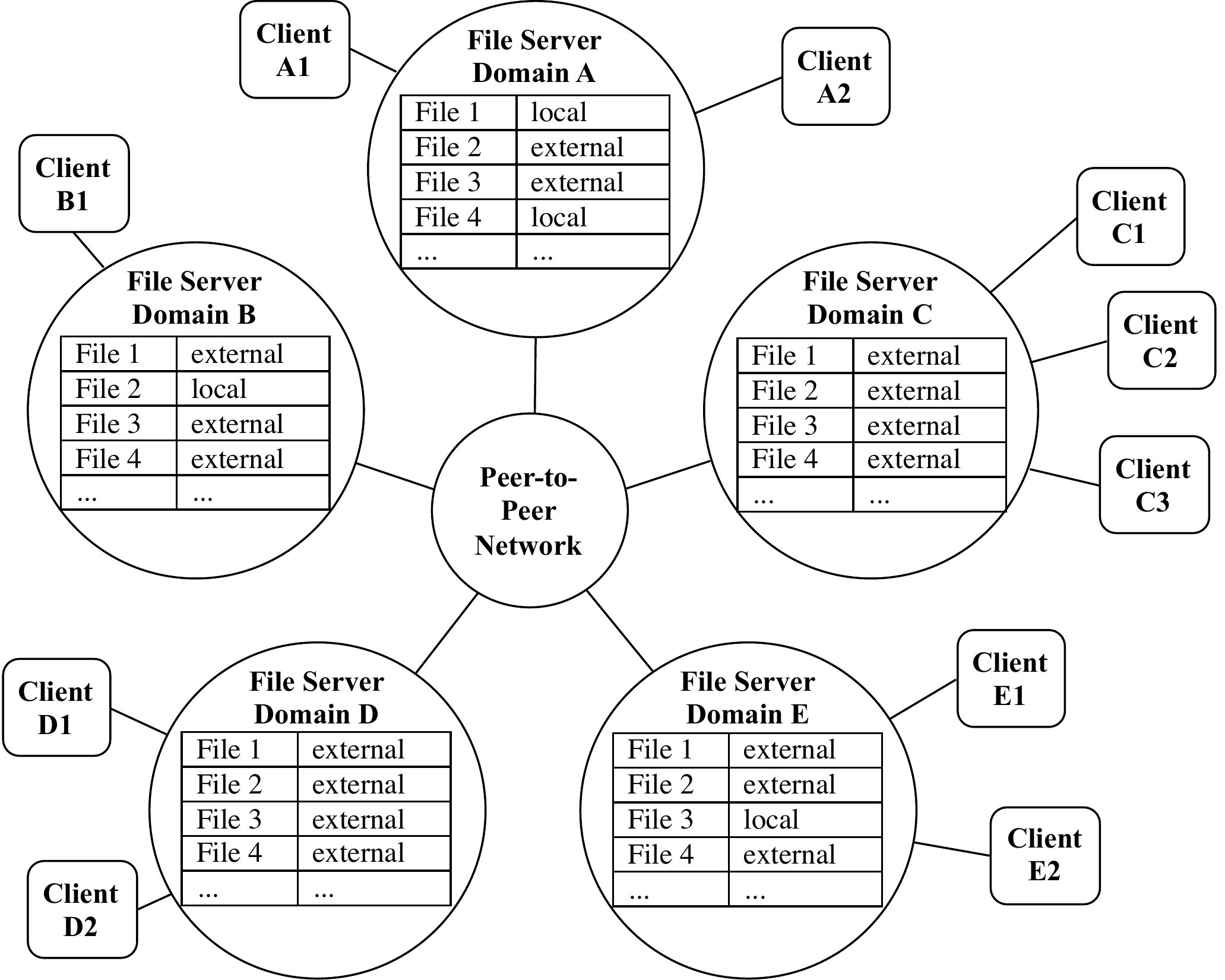}
    \caption{Transparent file server in a shared data pool.}
    \label{fig:transparentfileserver}
\end{figure}

P2P file distribution and download is usually performed by software that is installed at the user's machine or device, e.g. a torrent client or an IPFS node \cite{benet2014ipfs}, which acts as a peer. In the use-cases that are discussed in this article, peers only act on behalf of an organisation or institution that provides a number of files to a shared data pool. In such a setup, the P2P network is not created between the users of the data, but between the organisations or institutions only. As illustrated in figure \vref{fig:transparentfileserver}, each partner provides exactly one node to the P2P file-exchange network and users do not have direct access to that P2P network. 

Each node acts as a transparent file server to the users, showing all available data that is accessible in the shared data pool. A client that is accessing a file will not know whether the data is stored in the local file server (local) or whether it has to be downloaded from the P2P network first on request (external). After the file has been downloaded and decrypted, it is up to the local server to decide to keep it for future use and when the local copy would be deleted (e.g., due to space limitations).

In the described use case of a pool of research data the end-users are the researchers who access the genomics data in the local file server at the research institute. The research institute maintains a file server as a node in the shared data pool, which requests the needed data on demand or uploads new files that it receives from researchers.

\begin{figure}[tb]
    \centering
    \includegraphics[width=0.7\textwidth]{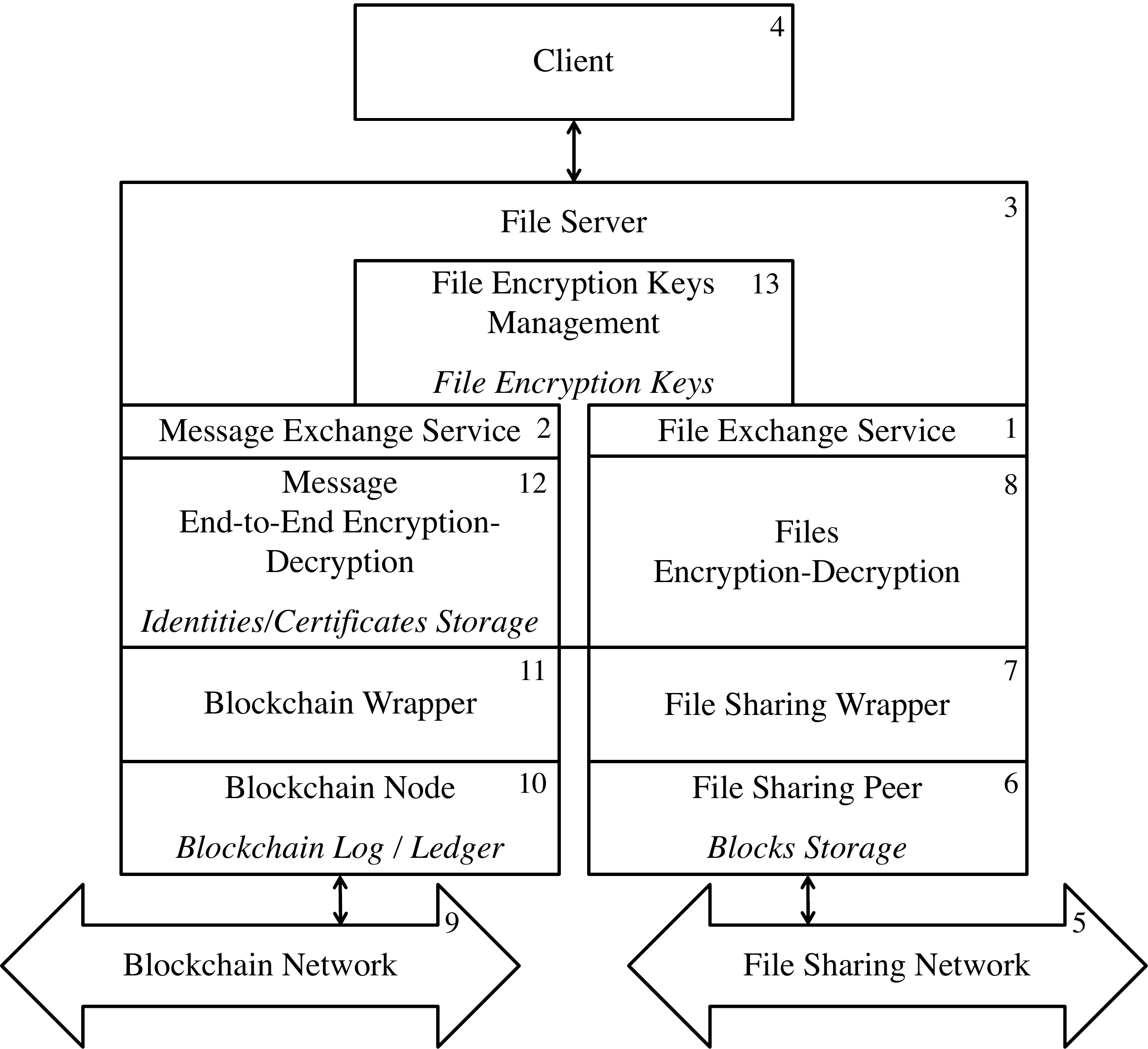}
    \caption{Main elements of the transparent file servers network.}
    \label{fig:innerstructure}
\end{figure}

Figure \vref{fig:innerstructure} shows the elements of the server and its used networks. The small numbers in the figure are referenced in the following description as superscript numbers. 

The server consists of three parts: A file exchange service that bases on a file sharing network;\textsuperscript{1} A message exchange service to exchange  administrative information that is based on a blockchain network.\textsuperscript{2} All relevant information about the availability of files, requests and provisioning of encryption keys are distributed via this messaging exchange feature; a file server service\textsuperscript{3} on top of all this that provides transparent access to clients\textsuperscript{4} to all files that are managed in the shared pool.

Each partner of the data pool that manages a file server, manages exactly one node in the file sharing network and exactly one node in the blockchain network. The file server needs to maintain the nodes in both networks synchronised and a failure of the file server affects always both networks.

Contrary to public blockchain or globally used file sharing networks, it is assumed, that the topology of both networks stays quite stable and only planned topology changes may occur. With that knowledge, the temporary downtime of a blockchain or file-sharing node must not lead to a restructure of the underlying network topology. This is quite important, globally for file sharing network to avoid unnecessary network traffic but also locally inside the server to keep the blockchain nodes and the file-sharing nodes synchronised.

\subsection{Message exchange}
\label{sec:messageexchange}
In the proposed concept (still figure \vref{fig:innerstructure}), the blockchain network\textsuperscript{9} is mainly used as a persistent blockchain based message exchange platform with a message exchange service (as shown in \cite{Roth2018a}).\textsuperscript{2} Each message that is sent by a blockchain node\textsuperscript{10} via the network is transparently visible for all the other participants and can never be changed. This is important, because this feature is used to introduce non-repudiation of the request for the unique file decryption key as described later in sub-section \ref{sec:filedecryption}. The request and the related responses are seen as a proof that the requester had plain-text access to the file that is linked to the requested key.

The underlying blockchain solution is not of importance, as long as it is possible to embed some bits and bytes inside a transaction to allow the submission of a message of random size. A blockchain specific wrapper layer\textsuperscript{11} is introduced to hide blockchain specific details like smart contracts or blockchain node identities from the higher layers of functionality and to ensure independence from the underlying blockchain solution. This is important to allow the replacement of a specific blockchain solution (e.g. due to licensing issues or scalability issues) without affecting the higher functionality. Only the wrapper layer has to be re-implemented and the provided functionality has to be translated into blockchain specific transactions.

As an example the way a message has to be split into segments and embed into several blockchain transactions (e.g. as part of an unspent transaction in Bitcoin\cite{Bradbury2013}) and the way these fractions have to be extracted out of the blockchain varies from the used blockchain solution. The wrapper layer can be seen as a driver for the specific blockchain that has to implement a predefined interface.

Messages that are sent via the blockchain can be sent as broadcast in plain-text to all participants in the network or end-to-end-encrypted\textsuperscript{12} to a specific number of recipients. The latter is done in an S/MIME style, so each message is first signed and then encrypted with a unique symmetric key that is individually encrypted for all recipients with their public key. This allowing every recipient that is in the list to decrypt the symmetric key on base of their own private key.

The following example from \cite{Roth2018a} (Listing \vref{lst:xmlstructure}) shows how all information is packed inside an XML structure. In the example the payload message is sent to two recipients. The symmetric key is not only encrypted for all recipients but also for the sender to allow it to read its own message. In case there are no receivers listed in that data structure, the payload will only consist of a signed pain-text message that is broadcast to all participants.

\begin{lstlisting}[float,floatplacement=H,frame=single,caption={XML structure of the message},captionpos=b,label={lst:xmlstructure}]
<?xml version="1.0" encoding="UTF-8" standalone="yes"?>
<MessageStructure>
  <Header>
    <Version> 1.0.0 </Version>
    <Sender>
      <ID Key="Base64(Encrypt(public key of sender, symmetric key))">
        Base64(identifier of the sender)
      </ID>
    </Sender>
    <Receiver>
      <ID Key="Base64(Encrypt(public key of recipient 1, symmetric key))">
       Base64(identifier of the recipient 1)
      </ID>
      <ID Key="Base64(Encrypt(public key of recipient 2, symmetric key))">
        Base64(identifier of the recipient 2)
      </ID>
    </Receiver>
  </Header>
  <Payload>
    Base64(SignThenEncrypt(private key of sender, symmetric key, message)
  </Payload>
</MessageStructure>
\end{lstlisting}
The message that is packed inside the payload section is a data structure that is defined by the higher layers. As examples, the message could be a request for a specific key, distribution of key shares or information about new files.

All recipients and the sender have unique identifiers which are different from their blockchain identity. This is important because the underlying blockchain solution is decoupled from the higher layers via the wrapper layer. This new type of identifier identifies the message endpoints but is different from their public keys. A later replacement of a public/private key pair can be done without affecting the unique identifier.

During the initial setup of the server, its identity has to be created, which leads to the creation of a certificate and a public/private key-pair. In case of the joining to an existing data pool not only IP-addresses and ports need mutually to be exchanged and accepted but also the identities and certificates of the servers. These certificates do not have to be signed by a known and accepted certification authority, because the pool is created as a closed group of known partners. The management and exchange of these certificates and keys is not in the focus of this article.

With the use of identities outside the blockchain, the sending of a message to a dedicated partner does not require a transaction of assets (coins) to the blockchain node of the receiver partner. In fact, to send the segments of a message, the transactions that carries the segments can transfer only minimal assets or coins between wallets that are all owned by the sender. Due to the transparency of the blockchain, other partners will still be able to read these transactions.

\begin{figure}[tb]
    \centering
    \includegraphics[width=0.8\textwidth]{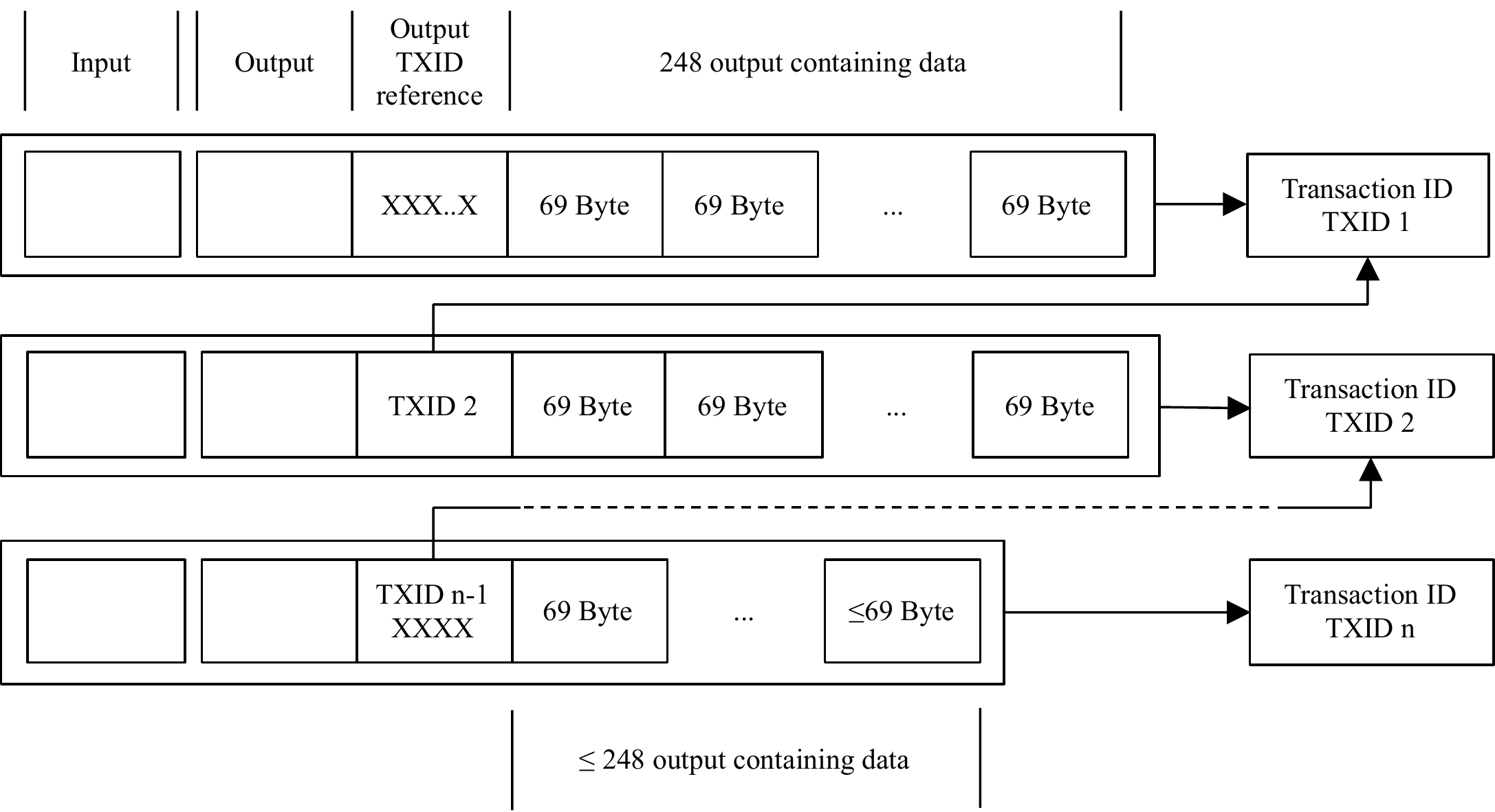}
    \caption{Segmentation of a message into blockchain transactions. \cite{Roth2018a}}
    \label{fig:segmentation}
\end{figure}

The XML structure as it is shown in the previous example is sent to the blockchain wrapper to distribute it via the blockchain. To give an idea how that wrapper is taking care of it, the process is explained for the wrapper that uses Bitcoin blockchain (see figure \vref{fig:segmentation}). 

A transaction in Bitcoin consists of inputs and outputs, which describe the movement of coins between the owner of the coins and the new owner. The number of outputs inside a transaction is limited to 250 transactions and an output that is tagged as unspent might only contain up to 70 bytes of random data (with one byte reserved for the size of data). Therefore, the entire message must be split into segments of 69 bytes maximum. The blockchain only accepts a transaction as valid in case the input coins are entirely consumed in at least one output. A second unspent output is needed to create the link between transactions in case the number of segments of the message does not fit into one transaction. The first transaction is indicated with 64 times "X" in the second output. For each of the following transactions this output will contain the transaction ID of the previous transaction (TXID), which will establish a chain of transactions. The last transaction that is needed to cover the message will contain additionally four "X".

Even in case transactions are executed in the wrong order on the blockchain, the reader of the blockchain is able to identify all transactions that belong to one message and merge the segments to one completed XML structure. This can be handed over to the higher layers to identify if is one of the foreseen receivers. In case of yes, the payload message can be decrypted and the command that had been sent can be processed. 

\subsection{File exchange}
\label{sec:fileexchange}
The use of a file sharing network\textsuperscript{5} (figure \vref{fig:innerstructure}) is seen as the best way to distribute files redundantly among the other peers\textsuperscript{6} as part of a file sharing service\textsuperscript{1}. In principle several solutions could be used to fulfil the desired purpose, e.g., Distributed Hash Table solutions like Pastry \cite{10.1007/3-540-45518-3_18} or Chord \cite{Stoica2001a}. Some additional demands do not allow to use the chosen solution out-of-the box. In case the file sharing solution does not depend on the distribution of chunks of the file, files need to be split before handing the chunks over to the file distribution network. It is also foreseen, that the source of a file always maintains its original at its server. The file sharing network can take this into consideration during the parallel download of chunks. Finally, before handing a file over to the file sharing network, it must be encrypted\textsuperscript{8} with a unique symmetric key per file before it is split into chunks. That symmetric key is also needed to decrypt files that have been loaded and reconstructed from the file sharing network. The management of these keys is one of the crucial elements in the proposed concept.\textsuperscript{13} For that, the message exchange feature is used.

For all these demands, a wrapper layer\textsuperscript{7} is introduced that decouples the specific details of the file sharing solutions from the higher functionalities. This is similar to the wrapper layer of the blockchain\textsuperscript{11} and fulfils the same purposes: the wrapper layer acts as a driver for the file sharing solution and a change of the file sharing solution only affect the wrapper layer. 

\subsection{File encryption and distribution}
\label{sec:fileencryption}

\begin{figure}[!tb]
    \centering
    \includegraphics[width=0.45\textwidth]{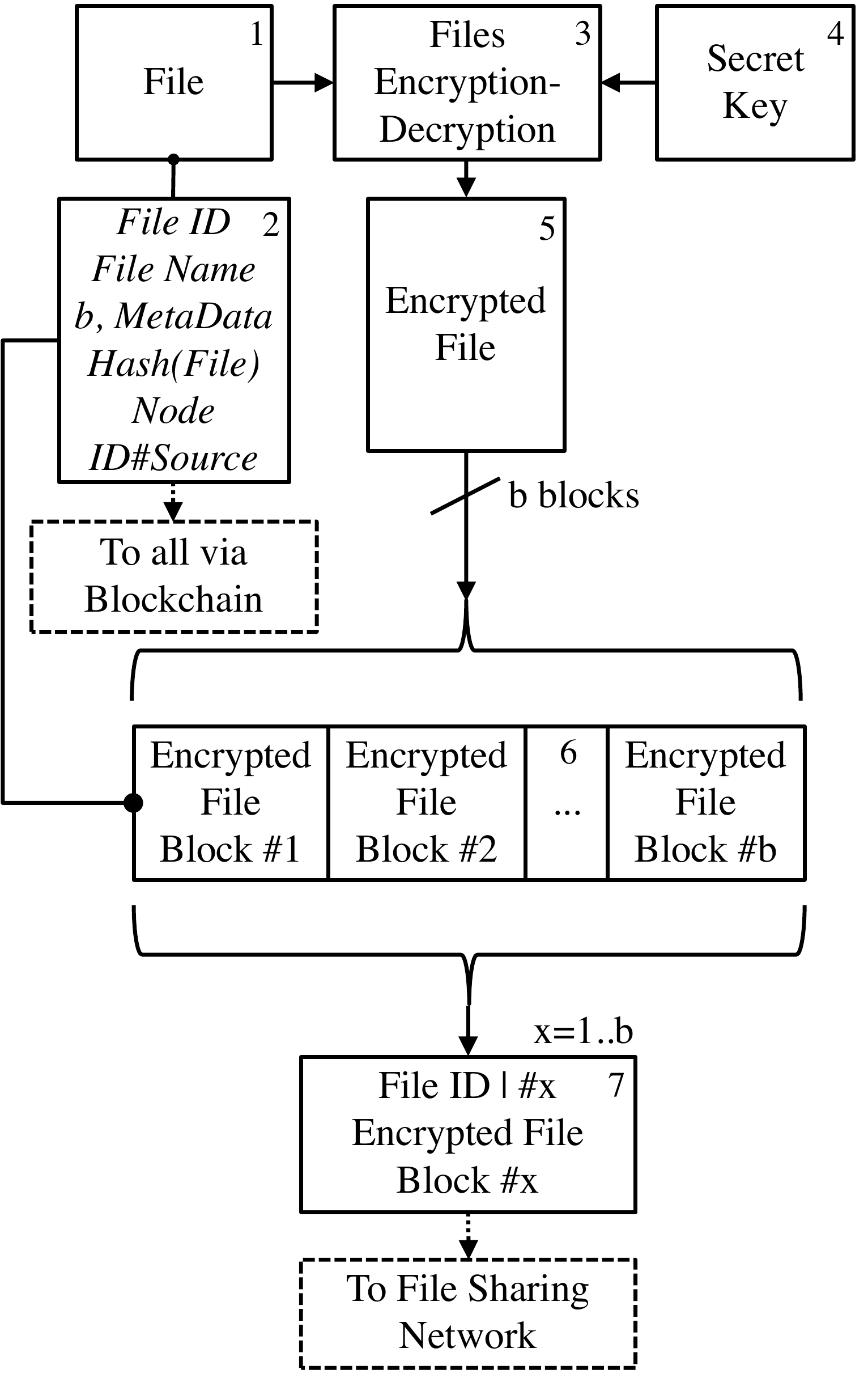}
    \caption{Process of file encryption and distribution}
    \label{fig:fileencryption}
\end{figure}

\begin{figure}[!hptb]
    \centering
    \includegraphics[width=0.6\textwidth]{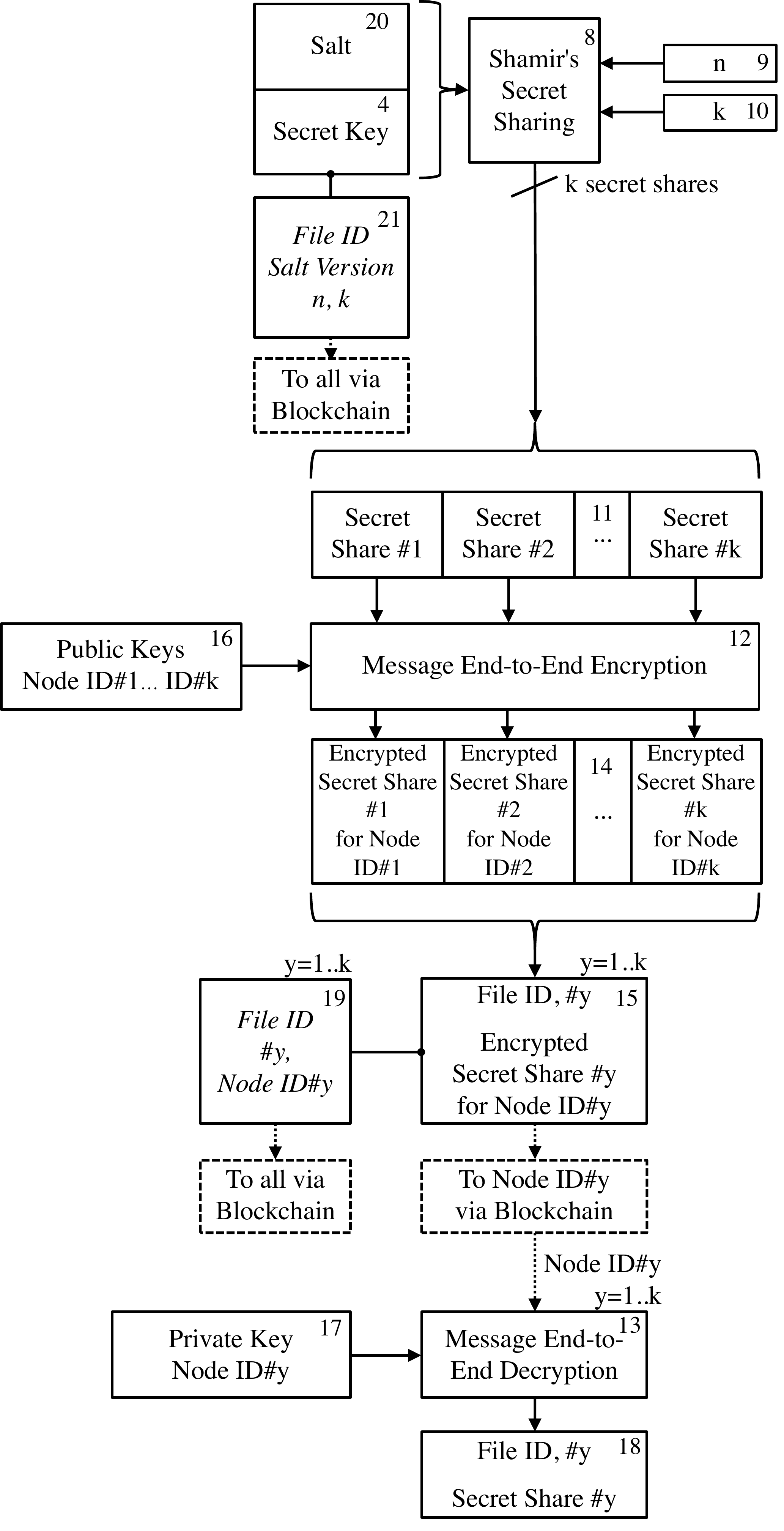}
    \caption{Process of distribution of secret shares}
    \label{fig:distributeKey}
\end{figure}

The process of file encryption and file distribution is drawn in figures \vref{fig:fileencryption} and figure \vref{fig:distributeKey}. Each file\textsuperscript{1} consist of a File ID, a File Name, the origin of the file (Node Identity of the Source) and some Meta Data\textsuperscript{2}. Before distributing the file over the file sharing network it has to be encrypted first\textsuperscript{3} with the use of a unique secret symmetric key\textsuperscript{4}. The encrypted file\textsuperscript{5} is then split into blocks of equal size (the last block might be smaller)\textsuperscript{6}. Since the initial file size is random, the number (\textit{b}) of blocks of a file will vary. Each block with number \textit{x} then is handed over to the File Sharing Network\textsuperscript{7}. The submission key that is used to identify that block is the File ID concatenated with the number of the block \textit{x}. To allow a retrieval of the file at other nodes, the File ID, the File Name (including path), the Node ID of the Source, the number of blocks \textit{b}, and the Meta Data of the file\textsuperscript{2} will be sent as a message to all in plain-text, using the blockchain network.

The secret key\textsuperscript{4} that was used for the encryption should not only resist on the source of the file, in case that server is temporary not available. To allow nodes to request the key, the key is distributed to all nodes in the network on base of secret shares (figure \vref{fig:distributeKey}). Shamir's Secret Sharing algorithm\textsuperscript{8} \cite{Shamir1979} is used to create \textit{k}\textsuperscript{10} secret shares for which \textit{n}\textsuperscript{9} out of these shares are needed to reconstruct the initial key. In the proposed solution, \textit{k} is the number of nodes in the data pool, and \textit{n} is set to $\lceil\sfrac{2}{3}\textit{k}\rceil$. That value was chosen to require a quite high number of malicious nodes who have to collaborate to retrieve the key without conforming the protocol.

Each secret share with a certain number\textsuperscript{11} is sent end-to-end encrypted\textsuperscript{12-15} by the use of public and private keys\textsuperscript{16,17} via the blockchain network to its assigned node, that stores the secret share together with the file number and the number of the secret share\textsuperscript{18}. A list of File ID, share number and responsible node is distributed to all, to allow an easy look-up of the secret shares\textsuperscript{19} and its assigned node.

\subsection{Versions of the secret shares}
\label{sec:secretshareversions}
As seen in figure \vref{fig:distributeKey}, the secret shares are not only created out of the secret key but also from a salt\textsuperscript{20}. This is important to allow versions of secret shares on base of different \textit{(n, k)} values due to potential topology changes. If the number of active nodes drops below \textit{n}, a reconstruction of the secret key is not possible anymore. In case that happens, it is always possible that the source creates a new set of secret shares on base of the used secret key, but with new \textit{n'} and \textit{k'} values. Depending on the secret sharing algorithm, these new secret shares might be compatible with the old ones, so it might be possible that a node could collect several secret shares from different versions which could allow the node to reconstruct the secret key. With the use of a version-depending salted secret key, this can be avoided. The less good alternative would have been to drop all encrypted file with insufficient secret shares from the file sharing network and re-send all these files encrypted with an entirely new encryption key. The set of File ID, the \textit{(n, k)} scheme that was used to split the encryption key of the file, and the current version of the salt is distributed to all via the blockchain\textsuperscript{21}.

Versions of the secret shares might also be used to invalidate files, by distributing a new version of pseudo secret shares, that cannot be used to reconstruct the secret key. In all cases, only the source is able to update secret shares of files and everyone is able to identify that by checking whether the signature of the message matches with the identity of the source of the file and its secret shares.

\subsection{File download and decryption}
\label{sec:filedecryption}

\begin{figure}[!hptb]
    \centering
    \includegraphics[width=0.5\textwidth]{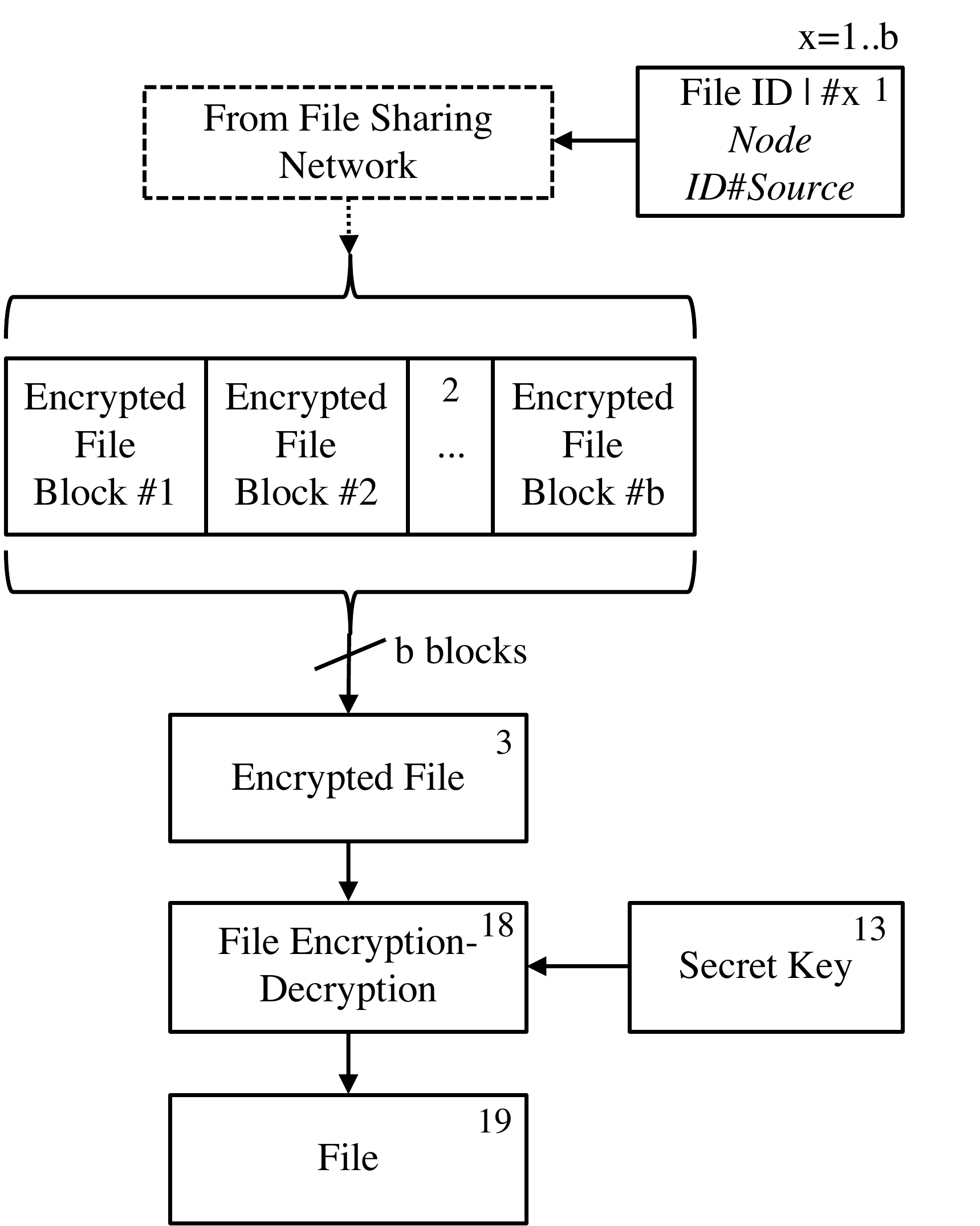}
    \caption{Process of file decryption}
    \label{fig:filedecryption}
\end{figure}

\begin{figure}[!hptb]
    \centering
    \includegraphics[width=0.65\textwidth]{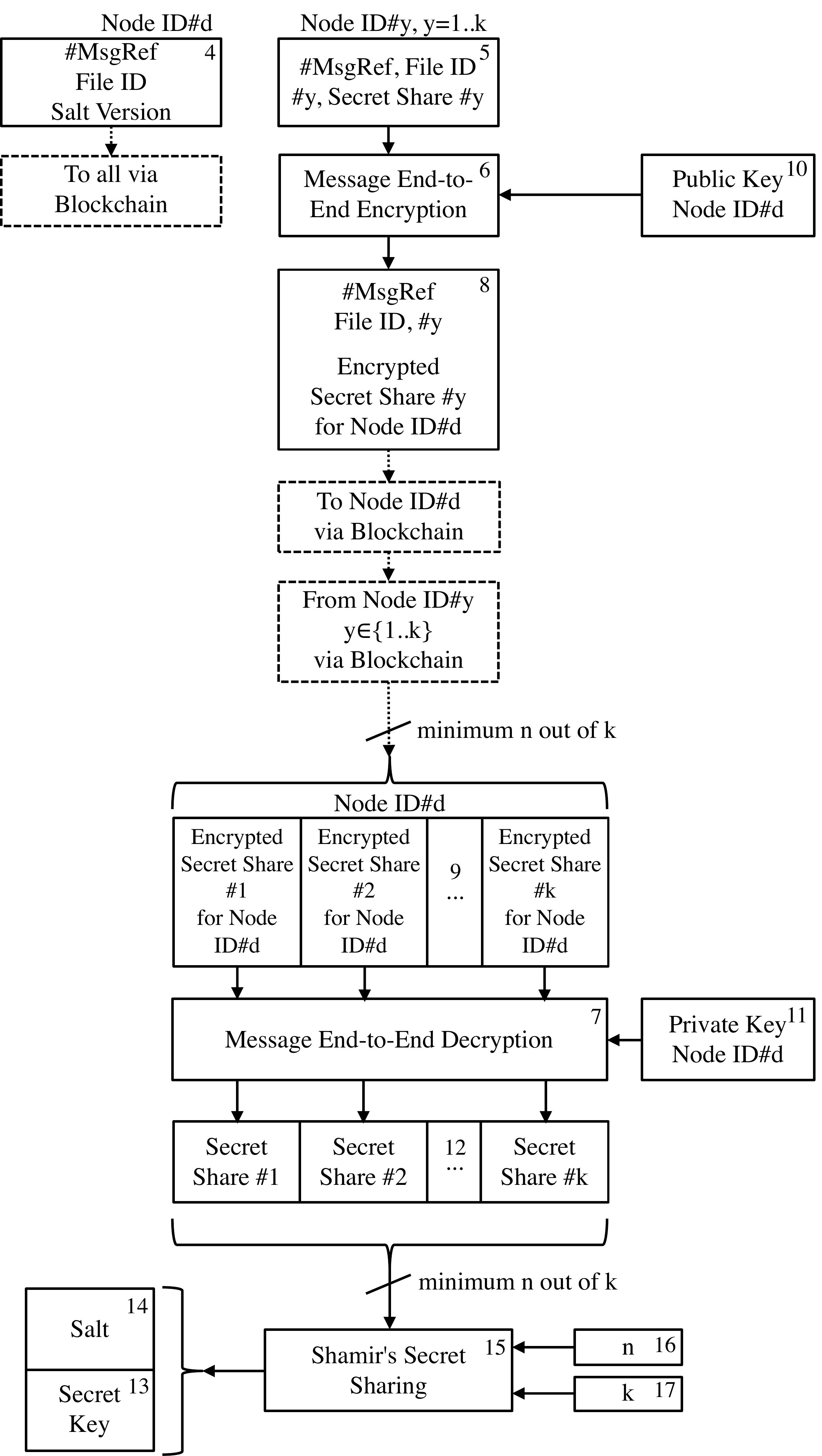}
    \caption{Process of shares retrieval}
    \label{fig:collectkey}
\end{figure}

As shown in figure \vref{fig:filedecryption}, with the use of the File ID and the maximal number of blocks of a certain file\textsuperscript{1}, a node is able to request all relevant blocks\textsuperscript{2} from the file sharing network to reconstruct the encrypted file\textsuperscript{3}. In worst cases, blocks also can be requested from the source, of the files.

The request of the blocks as well as the response cannot be seen as a proof for the access of the file because every communication can be denied by the node to have happened. Additionally, a node will at that stage always be able to claim to only have access to the encrypted file\textsuperscript{3}. For that the secret key {\small(13)} must be known to decrypt\textsuperscript{18} the file\textsuperscript{19}. This requires access to the minimum number of the secret shares that are needed to reconstruct the secret key.

As shown in figure \vref{fig:collectkey}, to retrieve a minimum number of \textit{n} out of \textit{k} secret share, a broadcast is sent to all nodes, requesting the secret shares for the File ID and the newest version of the used salt\textsuperscript{4}. This request is sent in plain-text with a message reference. This reference is used during the reply of the secret shares from the other nodes and is used to avoid that a node denies to have access to sufficient secret shares. Having sufficient replies of secret shares, it is seen as a proof, that a node had access to the secret key and so to the plain-text file.

All secret shares\textsuperscript{5} from the nodes are sent back end-to-end encrypted\textsuperscript{6-9} using the public\textsuperscript{10} and private key\textsuperscript{11} of the requesting node. If a node still denies to have access to the plain-text version of the file, the node has to prove, that the received messages containing the secret shares could either not be decoded on base of the nodes private key or that the decoded secret shares do not lead to a secret key that could decode the encrypted file.

Because all messages that are related to the secret shares as well as the hash value of the original file are transparent on the blockchain, it will always be possible to replay the reconstruction process and compare the outcome of that process with the claim. In that case it is a reversal of the burden of proof.

After the decryption of the minimum number of needed secret shares\textsuperscript{10,11}, the secret key\textsuperscript{13} and its salt\textsuperscript{14} are decrypted\textsuperscript{15-17}. With no surprise, the salt is not needed and is ignored. Finally the secret key is used to decrypt the file (figure \vref{fig:filedecryption}\textsuperscript{18-19}).

\section{Security}
\label{sec:security}
There are several attack surfaces that need to be analysed to estimate the possibilities of security problems: the used distributed hash table network, the blockchain network and the key management. Protocol specific problems cannot be discussed in that section, because the used blockchain or distributed hash table network is encapsulated by a wrapper layer and might change.

\subsection{Distributed hash table network}
\label{sec:distributedhashtablenetwork2}
All files in the distributed hash table network are stored encrypted. The concept assumes the possibility that all blocks of a file might accumulate at one node, so there is no need for a group of malicious nodes to cheat and exchange blocks outside the distributed hash table protocol.

The provisioning of false blocks on request also does not help, because the validity of a block can be checked by the use of its hash value.

It is possible not to respond to requests for blocks. With a certain number of malicious peers, it is possible, that for some of the blocks no response to requests follow, leaving the file incomplete. In that case it is still possible to request the blocks from the original source. This type of attack cannot be performed without being noticed: All non-responding peers can be identified. In case peers act in the described way more often, questions could be asked to the maintainers of the peer.

A malicious node could try to upload a huge amount of data to the peer-to-peer network, that are not linked to any file managed by the pool. In that case, the storage space of the partner peers will be occupied with useless data and might lead to a fully occupied storage space at some peers. This attack can be avoided, in case all peer-to-peer activity is synchronised with the use of administrative information that is distributed via the messaging service. Blocks that are uploaded without prior notification would then be dropped.

\subsection{Blockchain network}
\label{sec:blockchainnetwork2}
In case of 50\%+1 malicious node it is possible to revert the state of the blockchain and delete all transactions in a branch that have happened starting at a certain time. This attack could be used to delete the evidence, that a key was requested and that secret shares have been sent. The agreement of all nodes on the currently valid blockchain and its winning branch does not mean, that all nodes have to forget the loosing branch. Every node could keep that dropped branch and so still provide evidence of requesting the key in case of need.

\subsection{Key management}
\label{sec:keymanagement2}
There are \textit{n} out of \textit{k} secret shares needed to reconstruct the encryption key. \textit{n} is set to $\lceil\sfrac{2}{3}\textit{k}\rceil$ in the proposed solution. This means that there are roughly 66.6\% of malicious nodes needed who must exchange secret shares outside the protocol, before it is possible to reconstruct the key without leaving evidence in the blockchain. If such a high number of nodes are malicious, the data pool faces bigger problems than that.

With the use of a salted key, it is not possible to match newly created secret shares with secret shares that have already been distributed. This makes it impossible to collect versions of secret shares over time to reconstruct the key without the need to request other shares.

A node might distribute wrong secret shares to all nodes, which do not allow the decryption of a file. A node that has requested a file can provide proof that it was not able to decrypt the file: First it has to provide proof that its private key matches to the known public key that was used for the end-to-end encrypted message exchange. Secondly, it has to show that the decrypted secret shares do not lead to a valid key.

A similar attack is the provisioning of wrong secret share by only one or many malicious nodes in case of a request for secret shares. If the number of received secret shares \textit{n'} is bigger than minimum \textit{n}, a node is able to create subsets of \textit{n} shares out of the \textit{n'} received shares and reconstruct the secret key on base of that subset. All possible subsets that do not contain the wrong shares lead to the same correct key. With the identification of the wrong shares it is possible to identify the malicious nodes. Unfortunately, the number of wrong responses in the entire set of responses \textit{n'}, must only be greater than \textit{n'-n-1} to make the reconstruction impossible for all subsets because in each subset at least one malicious node will invalidate the reconstruction of the key. This becomes important in case an attacker is able to exclude nodes from responding and so to lower the number of total responses \textit{n'}. And having \textit{n} set to $\lceil\sfrac{2}{3}\textit{k}\rceil$, only approximately 33,3\% malicious nodes are needed to invalidate all possible subset of shares and making it impossible to reconstruct the key.

An easy way to address the problem of providing wrong secret share is to broadcast the hash value of each secret share. This has to be done by the provider of the secret shares and in plain-text. In that case, the receiver can check the validity of each received share by comparing their hash value with the broadcast version. With that, all malicious nodes can be identified and actions can be taken.

\section{Conclusions}
\label{sec:conclusion}
The proposed concept provides a solution for a specific setup, in which a pool of files is setup by a consortium without any access restriction to the files, but in which its first access by a partner has to be logged. Additionally, it is demanded, that the concept works without involvement of the source of a file or a central server and should not require special software or plug-ins at the end-user side.

Having that said, it might become clear, that the concept that introduces the blockchain as the tool to log the file access, will easily become obsolete in case some requirements are removed. As an example, if the symmetric key has to be downloaded from the source of the file or from a central server, a blockchain could be avoided, because the request of the key could be easily logged there.

Also, the number of real-life use cases that only demand the logging of the very first access of file that also can be accessed without further authorisation, might be limited. In case permission has to be asked, and access can be rejected, even in the future, the proposed setup is not the right choice. It would be even difficult to withdraw access permission, because once a file has been downloaded it will stay locally in plain-text, even if the key-access would be refused. But this is one of the big assets of the solution, because it does not require specialised access software or plug-ins that enforces the access permission.

Apart from that, the proposed solutions provide some concepts that can be re-used in other context. First, the use of the blockchain as a message exchange platform to distribute remote function execution requests has been discussed in detail in \cite{Roth2017,Roth2018a}. Secondly, the concept of secret shares on base of a salted key gives the opportunity to re-distribute new sets of secret shares of the same key, without the possibility to accumulate old shares and disclose the secret.



\bibliographystyle{IEEEtran}

\bibliography{references}

\end{document}